# Exploring the structural and optoelectronic properties of natural insulating phlogopite in van der Waals heterostructures

Alisson R. Cadore[1]*, Raphaela de Oliveira[2], Raphael L. M. Lobato[3], Verônica de C. Teixeira[4], Danilo A. Nagaoka[1,5], Vinicius T. Alvarenga,[1,5] Jenaina Ribeiro-Soares[3], Kenji Watanabe[6], Takashi Taniguchi[7], Roberto M. Paniago[2], Angelo Malachias[2], Klaus Krambrock[2], Ingrid D. Barcelos[4], Christiano J. S. de Matos[1,5]

[1]*School of Engineering, Mackenzie Presbyterian University, São Paulo, Brazil*

[2]*Physics Department, Federal University of Minas Gerais, Minas Gerais, Brazil*

[3]*Physics Department, Institute of Natural Science, Federal University of Lavras, Minas Gerais, Brazil*

[4]*Brazilian Synchrotron Light Laboratory (LNLS), Brazilian Center for Research in Energy and Materials (CNPEM), São Paulo, Brazil*

[5]*MackGraphe, Mackenzie Presbyterian University, São Paulo, Brazil*

[6]*Research Center for Functional Materials, National Institute for Materials Science, Tsukuba, Japan*

[7]*International Center for Materials Nanoarchitectonics, National Institute for Materials Science, Tsukuba, Japan*

*Corresponding author: alisson.cadore@mackenzie.br*

Naturally occurring van der Waals crystals have brought unprecedented interest to nanomaterial researchers in recent years. So far, more than 1800 layered materials (LMs) have been identified but only a few insulating and naturally occurring LMs were deeply investigated. Phyllosilicate minerals, which are a class of natural and abundant LMs, have been recently considered as a low-cost source of insulating nanomaterials. Within this family an almost barely explored material emerges: phlogopite [KMg$_3$(AlSi$_3$)O$_{10}$(OH)$_2$]. Here we carry out a high throughput characterization of this LM by employing several experimental techniques, corroborating the major findings with first-principles calculations. We show that monolayers (1L) and few-layers of this material are air and temperature stable, as well as easily obtained by the standard mechanical exfoliation technique, have an atomically flat surface, and lower bandgap than its bulk counterpart, an unusual trend in LMs. We also systematically study the basic properties of ultrathin phlogopite and demonstrate that natural phlogopite presents iron impurities in its crystal lattice, which decreases its bandgap from about 7 eV to 3.6 eV. Finally, we combine phlogopite crystals with 1L-WS$_2$ in ultrathin van der Waals heterostructures and present a photoluminescence study, revealing a significant enhancement on the 1L-WS$_2$ optical quality (*i.e.,* higher recombination efficiency through neutral excitons) similarly to that obtained on 1L-WS$_2$/hBN heterostructures. Our proof-of-concept study shows that phlogopite should be regarded as a good and promising candidate for LM-based applications as a low-cost layered nanomaterial.

## 1- Introduction

Layered materials (LMs)[1], for which graphene is the prime example, have obtained vast attention in the last years. So far, more than 1800 LMs have been computationally identified[2–4], while many of them have been naturally obtained or synthesized[5–13]. Thus, as soon as a new LM is identified, the investigation of its optical, mechanical and electrical properties is promptly examined[14]. Moreover, with the advent of techniques able to stack LMs precisely one on top of another creating the so-called van der Waals heterostructures (vdWHs)[15], new applications and studies are envisioned[16,17]. Consequently, individual LMs and their vdWHs are often considered as building blocks for future optoelectronic devices[14–17].

Despite the massive use of synthetic hexagonal boron nitride (hBN) as an insulator in vdWHs, the search for earth-abundant LMs sources that could potentially replace hBN is of great importance[18–20]. In this scenario, the mica group, which is one of the groups of sheet silicate (phyllosilicate) minerals and shows perfect basal cleavage[21,22], has driven the interest of many research groups because of its thermal, mechanical and electrical performance[23–28], its easy assembly in vdWHs[29,30], and its use as substrate in LMs growth[31,32]. All these qualities have boosted its applications[33]. Nevertheless, most of the investigations have focused on muscovite mica, and barely exploited other minerals of this family, for instance, phlogopite. Phlogopite is a common phyllosilicate within the mica group with chemical formula[34] $KMg_3(AlSi_3)O_{10}(OH)_2$. Some works have been done regarding its mechanical[35] and structural stability[36,37], electrical[27,28] and optical[38–41] properties in the bulk form, and its liquid phase[42] and mechanical[5,25] exfoliation capability. However, to the best of our knowledge, no systematical study has been performed to investigate the basic properties of phlogopite in the ultrathin form and its use in vdWHs with other LMs.

Here, we demonstrate that phlogopite is air-stable, can be mechanically exfoliated down to monolayers (1L), and has atomically flat surfaces like those obtained in hBN flakes. We present a robust experimental characterization of the phlogopite crystals by means of X-ray diffraction (XRD), X-ray Fluorescence mapping (XRF), wavelength-dispersive spectroscopy (WDS), electron paramagnetic resonance (EPR), optical absorption measurements (UV-Vis-NIR), atomic force microscopy (AFM), Raman, Mössbauer and Fourier-transform infrared (FTIR) spectroscopies, in addition with first-principles calculations. We then describe the basic structural, electronic structure, and optical features of this crystal and demonstrate that natural phlogopite presents iron (Fe) impurities in different sites and valence states in its crystal lattice, which decrease its bandgap from about 7 eV to 3.6 eV mainly by the incorporation of $Fe^{2+}$ in $Mg^{2+}$ sites. We demonstrate that the thickness reduction also leads to a bandgap decrease, an unusual trend in LMs. Finally, we show that insulating phlogopite supports thermal and hydrogen annealing, and it can be easily embedded in vdWHs, increasing the photoluminescence (PL) quality of 1L-$WS_2$ (*i.e.,* higher recombination efficiency through neutral excitons, $X^0$) similarly to what is achieved in 1L-WS$_2$/hBN heterostructures. Therefore, we foresee the use of phlogopite layers in ultrathin electronics and flexible devices[43,44], as well as in experiments dealing with nanophotonics[45,46].

## 2 - Results and Discussion

We start by discussing the structural characterization of a natural phlogopite crystal. From the analysis, we obtain a complete structural description of the sample examined. Mineral phlogopite (inset in Figure 1a), obtained from Minas Gerais (Brazilian mine at Itabira city), was used to produce the exfoliated samples investigated in this work. To confirm the structure and chemical composition of the bulk material, we initially investigated it by XRD (see Methods for details). Figure 1a plots the XRD diffractogram retrieved from a powdered sample of the crystal using a conventional laboratory source and Cu K-alpha radiation ($\lambda$ = 1.541Å). The sharp and intense peaks clearly show the crystalline nature of the sample. Given the mineral origin of the sample, different phases of phyllosilicates were searched to provide possible fits with the presence of major and minor phases. After going through search algorithms, Rietveld refinement was carried out using the MAUD package[47]. The best fitting to the experimental data, matching peak intensities and positions is retrieved for a volumetric combination of 98% phlogopite[48] and 2% biotite[49]. The parameters retrieved for the dominant phlogopite phase are $a$ = 5.338 Å, $b$ = 9.229 Å, $c$ = 10.284 Å, $\alpha$ = 90°, $\beta$ = 99.993°, and $\gamma$ = 90°. A fit of the peaks using only the phlogopite phase is shown in Figure 1a. The retrieved lattice values are consistent with those reported in the literature[50,51], but the lattice parameters may vary with the increasing contents of cations with different ionic radii[52]. The mineral exhibits a lamellar structure oriented along the (001) direction. The pristine crystal display monoclinic lattice, with symmetries from the space group P$^2$1 (*C2/m*), where two tetrahedral silicate sheets are adjacent to one octahedral layer – TOT structure, also known as 2:1 layered silicate structure[21,22,53], and its crystallographic representation is given in Figure 1b. Pristine phlogopite ("colorless", Fe-free crystal) is a hydrated phyllosilicate of potassium (K) and magnesium (Mg) with chemical formula KMg$_3$AlSi$_3$O$_{10}$(OH)$_2$, while the Fe-endmember ("brown" mineral) is the annite mineral (KFe$_3$AlSi$_3$O$_{10}$(OH)$_2$), and the biotite mineral can be seen as a Fe-rich crystal with chemical formula (K(Mg,Fe)$_3$AlSi$_3$O$_{10}$(OH)$_2$)[21,22,53]. However, in natural materials the existence of impurities such as Fe ions in their crystal lattice is reported in references[13,54–56]. Consequently, it is common to observe Fe:phlogopite crystals (light and dark brown minerals) with different Fe ions concentrations[34,38,56].

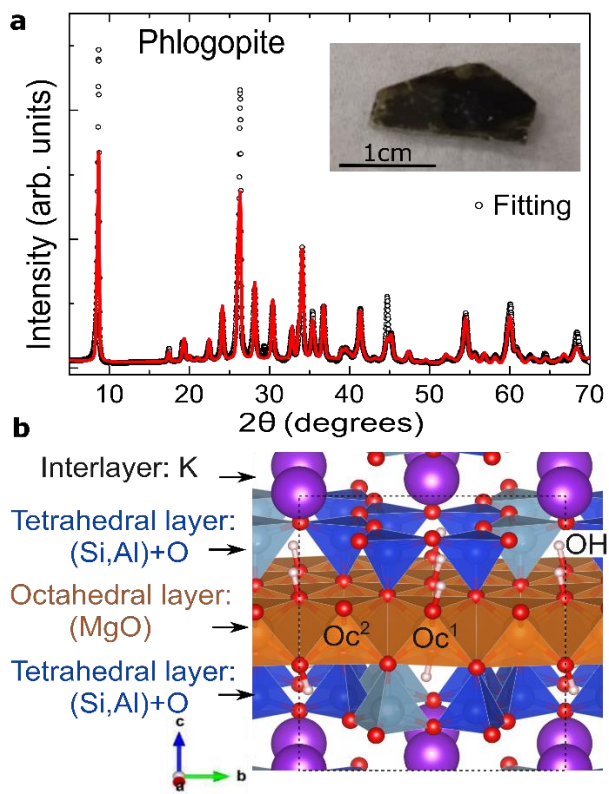

**Figure 1:** Phlogopite structure and its atomistic representation. a) A representative XRD diffractogram of the phlogopite crystal collected in a powder material. The inset shows photography of the crystal (mine rock) used in this work. b) Atomistic model of pristine phlogopite viewed from low-angle across the basal plane. The blue polygons are the tetrahedrons composed of Al, Si, and O, where Al and Si atoms exist at a ratio of 1:3. Two tetrahedral layers with their tetrahedral apex pointing towards each other sandwich an octahedral layer. The MgO octahedral units are indicated by light orange color. The interlayer region is occupied by K atoms. The Si and Al atoms are shown as dark and light blue spheres, respectively. The K atoms are shown as large purple spheres. The H and O atoms are denoted as white and red spheres, respectively. The separation between the K layers is approximately 1.0 nm, which is the thickness of mechanically exfoliated phlogopite. Fe ions at $Oc^1$ position are surrounded by four O ions and two OH ions at the corners, while at $Oc^2$ position the other octahedral site is characterized by adjacent OH ions. The small compass at the left bottom of the image indicates the disposition of the lattice vectors, and black-dashed lines indicate the supercell.

Next, the chemical composition of the phlogopite sample is determined by WDS, XRF mapping, EPR, and UV-Vis (see Methods for details regarding the measurements). Quantitative analysis by WDS provide a content of 41.9(8)% of $SiO_2$, 24.7(6)% of MgO, 15.6(3)% of $Al_2O_3$, 6.9(1)% of $K_2O$, 5.0(1)% of FeO and a content about 1% of other trace impurities, while the remaining contribution is due to the presence of water in the mineral structure. The WDS chemical evaluation indicates different impurities in natural phlogopite, with Fe being the predominant impurity. In accordance with this data, the XRF emission and mapping demonstrate qualitatively that Fe ions are largely predominant (Figure S1) and homogeneously distributed in the analyzed area as shown in Figure 2a, respectively. In the Supplementary Information, the Raman and FTIR analizes of our phlogopite sample also suggest the existence of impurities. Fe impurities are frequently present in the phlogopite structure as substitutional ions to $Al^{3+}$ or $Mg^{2+}$, being able to occupy both tetrahedral and octahedral sites[57]. The Fe ions can be present in the phlogopite structure with different valence states, $Fe^{2+}$ and $Fe^{3+}$. However, $Fe^{2+}$ ions with $3d^6$ electronic configuration are non-Kramer ions, being generally unobservable by EPR[58,59]. For the $Fe^{3+}$ ($3d^5$) ions with an effective electronic spin S = 5/2, at least four orthorhombic sites have already been identified by the EPR technique[57,60,61], being two tetrahedral and two (most commonly reported) octahedral sites. One octahedral site is determined by the Fe ion surrounded by four oxygen (O) ions and two hydroxyl (OH) ions at opposite vertices ($Oc^1$), while the other octahedral site ($Oc^2$) is characterized by OH ions occupying adjacent vertices[57,61]. The $Fe^{3+}$ ions experience a strong orthorhombic crystalline electric field that is sufficient to raise the degeneracy of Kramer's doublets with the environment symmetry varying from axial to rhombic[57]. As a result of the strong crystalline field, the octahedral substitutional sites for the Fe impurities must be distorted[57,60,61]. The octahedral site ($Oc^2$) with adjacent OH ions is the most rhombic one, resulting in an isotropic EPR line at $g \sim 4.2$ in phlogopite samples[57,60,61], while the octahedral site ($Oc^1$) with OH ions at the opposite vertices presents symmetry with nearly axial character[57]. EPR measurements at 75 K performed rotating the phlogopite sample with the $c$ axis perpendicular to the rotation axis (Figure 2b) reveal an approximately isotropic line at $g \sim 4.2$ that are characteristic to the most rhombic octahedral site ($Oc^2$) for substitutional Fe ions in phlogopite. Some broad anisotropic lines at lower magnetic fields are also observed, possibly due to the substitutional $Fe^{3+}$ ions in the $Oc^1$ site. However, due to the widening of the EPR lines, it is not possible to conclude about the presence or not of substitutional $Fe^{3+}$ impurities also in tetrahedral sites from our EPR spectra. For the angular dependence of EPR spectra with the rotation axis parallel to the $c$ axis (not shown), only a broad unresolved isotropic signal around $g \sim 4.2$ is observed. Our EPR measurements agree with previous works[41,57,60,61] and suggest that substitutional $Fe^{3+}$ ions are present in the phlogopite sample at least in one octahedral sites distorted by strong rhombic crystalline field. The broadening of the anisotropic lines in lower magnetic fields could be associated to the exchange interaction between the Fe impurities or due to the superposed slightly different orientations of the paramagnetic centers contributing to EPR spectra due to non-aligned stacking regions of the phlogopite lamellar structure.

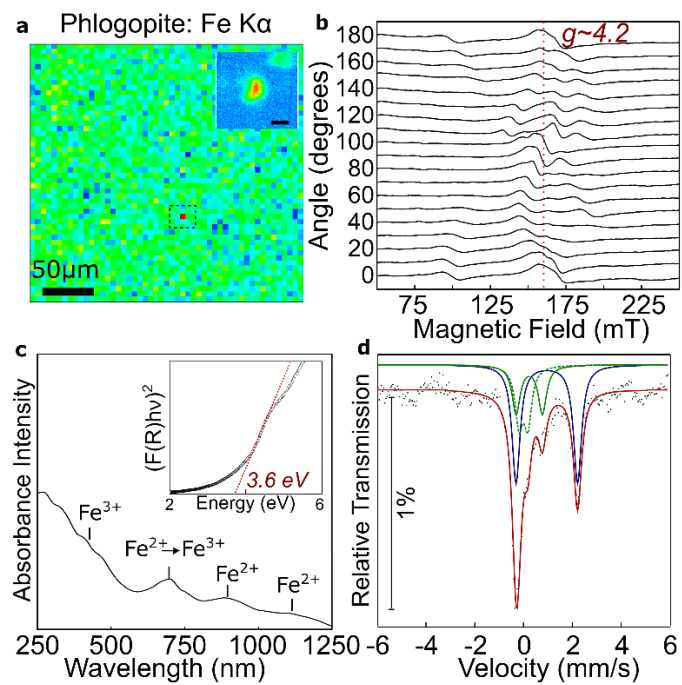

**Figure 2:** Experimental characterization of Fe:phlogopite sample. a) XRF mapping for Fe impurities over a 250 x 250 $\mu m^2$ region and over an isolated Fe island (inset, 5 x 5 $\mu m^2$, scale bar 1 $\mu m$) on phlogopite crystal. The images are obtained from the selection of Fe K alpha emission line, excited under 7.2 keV, and normalized by the intensity of the hottest pixel of each image. Red (blue) indicates a higher (lower) Fe concentration. b) The angular dependence of phlogopite EPR spectra at 75 K with 9.440(1) GHz performed with the *c* axis perpendicular to the rotation axis indicate a typical dependence of $Fe^{3+}$ ions in distorted octahedral sites. c) The absorbance spectrum of a phlogopite flake confirms the presence of Fe impurities in different valence states. The Tauc plot (inset) based on the reflectance measurement of a phlogopite flake indicates an experimental direct optical bandgap of about 3.6 eV determined by the Kubelka-Munk method. d) Fitted (red line) Mössbauer phlogopite spectrum indicating the presence of $Fe^{2+}$ (blue line) at octahedral sites and $Fe^{3+}$ at octahedral (continuous green line) and tetrahedral (dashed green line) sites with contributions of 53.7% and 46.3%, respectively.

Since EPR measurements are not sensitive to $Fe^{2+}$ ions, UV-Vis-NIR absorption measurements and Mössbauer spectroscopy are then performed (see Methods for details regarding the measurements). The absorbance spectrum of our phlogopite sample (Figure 2c) reveals bands associated with Fe impurities in both $Fe^{2+}$ and $Fe^{3+}$ valence states. The various absorption bands observed below 600 nm are due to orbital electronic transitions of $Fe^{3+}$ ions, while the two broad bands at approximately 900 nm and 1100 nm are due to the presence of $Fe^{2+}$ ions also in octahedral sites[62,63]. The absorption band observed around 700 nm is due to the charge transfer between the Fe ions[62,63]. Now, in order to determine the experimental optical bandgap of our phlogopite sample, a Tauc plot is elaborated from reflectance measurements of a phlogopite flake (inset in Figure 2c). The bandgap is obtained by extrapolation of the linear behavior (red dashed line) of the Kubelka-Munk function of the data[64,65]. Considering a direct bandgap[41] for phlogopite minerals, we estimate an experimental bandgap of about 3.6 eV for our sample that is in agreement with previous work on natural phlogopite[66], and lower than that in synthetic phlogopite[67] of 7.8 eV. The fitted phlogopite Mössbauer spectrum is shown in Figure 2d. The spectrum is fitted (red line) with three contributions of Fe doublets: the blue line component is assigned to $(VI)Fe^{2+}$ doublet and it corresponds to bivalent Fe impurities in octahedral sites (six-coordination), while the green lines components are assigned to $(VI)Fe^{3+}$ (continuous line) and $(IV)Fe^{3+}$ (dashed line) doublets, corresponding respectively to trivalent Fe impurities in octahedral and tetrahedral sites (four-coordination). The retrieved hyperfine parameters from the fit are 1.11, 0.39 and 0.16 mm/s for the isomeric shift and 2.52, 1.08 and 0.37 mm/s for the quadrupole splitting for the $(VI)Fe^{2+}$, $(VI)Fe^{3+}$ and $(IV)Fe^{3+}$ doublets, respectively. Quantifying the relative contribution of Fe ions from the absorption areas, we obtain a contribution of 53.7% due to $Fe^{2+}$ ions and 46.3% to $Fe^{3+}$ ions (22.0% in octahedral sites and 24.3% in tetrahedral sites). Thus, we can conclude that the incorporation of Fe impurities in the phlogopite structure occurs preferentially in the bivalent state at octahedral sites, while the $Fe^{3+}$ ions are present as substitutional impurities in both octahedral and tetrahedral sites, assuming different coordinations for the trivalent Fe ions in agreement with the literature[62,68,69]. This result complements our previous analysis by XRF, EPR and absorption measurements, giving a complete experimental description of the Fe incorporation in the phlogopite mineral.

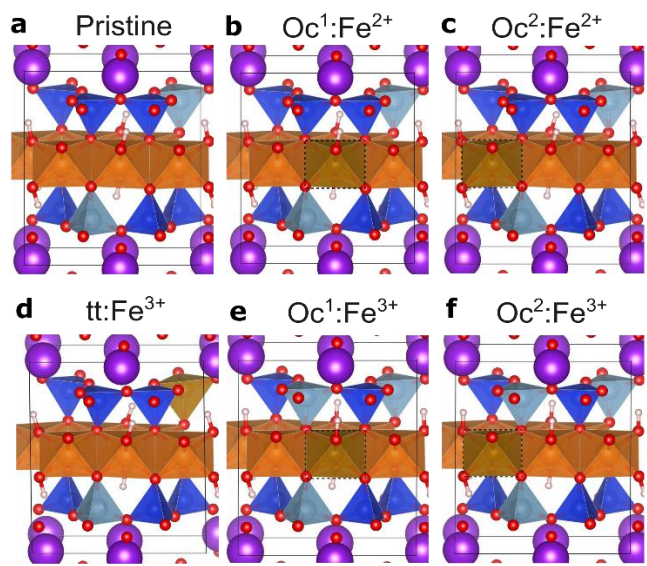

**Figure 3:** Atomistic model of the analysed Fe:phlogopite structures. a) Pristine phlogopite ($Mg_6(Al_2Si_6)O_{20}(OH)_4K_2$). Phlogopite ($Mg_5Fe(Al_2Si_6)O_{20}(OH)_4K_2$) with $Fe^{2+}$ ions at the (b) $Oc^1$ and (c) $Oc^2$ sites, respectively. d) Phlogopite ($Mg_6(AlFeSi_6)O_{20}(OH)_4K_2$) with $Fe^{3+}$ ions at the tetrahedral (tt) site. Phlogopite ($Mg_5Fe(Al_3Si_5)O_{20}(OH)_4K_2$) with $Fe^{3+}$ ions at the (e) $Oc^1$ and (f) $Oc^2$ sites, respectively.

The presence of impurities in micas were pointed as a possible reason for the lower bandgap of natural micas in comparison to that of synthetic micas[41,67]. In order to understand the influence of Fe impurities on the electronic structure of phlogopite, we perform density functional theory (DFT) calculations (see Methods for more details), considering the five most likely substitutional sites for Fe ions. These sites are based on our experimental results, previous literature[62,68,69], and the Hume-Rothery rule[70]. Figure 3a shows the supercell atomistic model for the pristine phlogopite ($Mg_6(Al_2Si_6)O_{20}(OH)_4K_2$). Figure 3b-c illustrates the isovalent substitutions of $Mg^{2+}$ by $Fe^{2+}$, where a single $Fe^{2+}$ ion occupies the two inequivalent octahedral sites $Oc^1$ and $Oc^2$, respectively. The substitution of $(VI)Fe^{2+}$ at the $(VI)Mg^{2+}$ site results in a difference of effective ionic radii, $\delta r$, of approximately 8%. Figure 3d shows the case for an isovalent substitution of $Al^{3+}$ by $Fe^{3+}$ ion, occupying the tetrahedral site of Al, named as tt site. Although the substitution of $(IV)$ $Al^{3+}$ by $(IV)$ $Fe^{3+}$ violates the Hume-Rothery rule ($\delta r \sim 25\%$), it occurs in nature[62,68,69]. Other cases for $Fe^{2+}$ or $Fe^{3+}$ ions in the tetrahedral layer, as $(IV)Fe^{2+}$ at the $(IV)Al^{3+}$ site ($\delta r \sim 62\%$), $(IV)Fe^{2+}$ at the $(IV)Si^{4+}$ site ($\delta r \sim 142\%$), $(IV)Fe^{3+}$ at the $(IV)Si^{4+}$ site ($\delta r \sim 89\%$) were neglected, due to the large $\delta r$, and lack of experimental evidence in our data. Figure 3e-f shows the case of substitution of $Mg^{2+}$ by $Fe^{3+}$ ions in the $Oc^1$ and $Oc^2$, respectively. These are no isovalent substitutions, and we consider the additional substitution of $Si^{4+}$ by $Al^{3+}$ ion for charge compensation, since its substitution is naturally present in the phlogopite structure. The calculated lattice parameters of pristine phlogopite, by using local-density approximation for the exchange-correlation functional (LDA), are $a = 5.169$ Å, $b = 8.958$ Å, $c = 9.842$ Å, $\alpha = 90°$, $\beta = 100.6°$, and $\gamma = 90°$, in close agreement to our experimental findings. Table 1 summarizes the structural changes of phlogopite induced by the different Fe ion sites analyzed.

Table 1: Structural parameters of phlogopite crystals, as calculated by using LDA exchange-correlation functional. The lattice constants, $a,b,c$; volumes of the supercell, V(cell); volumes of the octahedral site $Oc^1$, $V(Oc^1)$ and $Oc^2$, $V(Oc^2)$, where the substitution of $Mg^{2+}$ by $Fe^{2+}$ or $Fe^{3+}$ ions were considered; volume of the tetrahedral site $T_{Si}$, $V(T_{Si})$, where the substitution of $Si^{4+}$ by $Al^{3+}$ ion, and volume of the tt site, V(tt), where the substitution of $Al^{3+}$ by $Fe^{3+}$ ion were considered, are in angstrom. The angles between the lattice constants, $\alpha$, $\beta$, and $\gamma$ are in degrees.

| | *a* | *b* | *C* | α | B | γ | V(cell) | $V(Oc^1)$ | $V(Oc^2)$ | $V(T_{Si})$ | V(tt) |
|---|---|---|---|---|---|---|---|---|---|---|---|
| Pristine | 5.169 | 8.958 | 9.843 | 90.00 | 100.56 | 90.00 | 447.99 | 11.039 | 11.037 | 2.160 | 2.641 |
| $Oc^1:Fe^{2+}$ | 5.183 | 8.999 | 9.858 | 90.01 | 100.66 | 89.96 | 451.89 | 12.207 | 10.985 | 2.164 | 2.650 |
| $Oc^2:Fe^{2+}$ | 5.186 | 8.989 | 9.852 | 89.91 | 100.43 | 89.90 | 451.64 | 10.993 | 12.268 | 2.161 | 2.654 |
| $Oc^1:Fe^{3+}$ | 5.195 | 9.005 | 9.850 | 89.78 | 100.41 | 89.95 | 453.23 | 11.035 | 11.110 | 2.620 | 2.638 |

| | | | | | | | | | | |
|---|---|---|---|---|---|---|---|---|---|---|
| $Oc^2$:$Fe^{3+}$ | 5.195 | 9.006 | 9.8446 | 89.73 | 100.43 | 89.97 | 453.11 | 11.109 | 11.018 | 2.626 | 2.632 |
| $tt$:$Fe^{3+}$ | 5.193 | 9.028 | 9.898 | 90.81 | 101.22 | 90.19 | 455.09 | 11.127 | 11.189 | 2.149 | 3.449 |

The presence of Fe ions increases the lattice parameters and cell volume V(cell), regardless the occupation site. The volume of the octahedral sites $V(Oc^1)$ and $V(Oc^2)$ are essentially equal in the pristine material. In the phlogopite with $Fe^{2+}$ ions, the volume of the octahedron site with the $Fe^{2+}$ ion is greater than that with $Mg^{2+}$. In more detail, for the $Oc^1$:$Fe^{2+}$ geometry, the $Fe^{2+}$ ion occupies the $Oc^1$ site, with a volume $V(Oc^1)$ of about 11% larger than that of the $Oc^2$ site, which host the $Mg^{2+}$ ion. A similar trend is obtained for the relative volumes of the $V(Oc^1)$ and $V(Oc^2)$ site in the $Oc^2$:$Fe^{2+}$ geometry, where $Fe^{2+}$ and $Mg^{2+}$ ions exchange places in comparison to the $Oc^1$:$Fe^{2+}$ geometry. In the case of $Fe^{3+}$ ions in the octahedra site, the volume of the octahedron hosting the $Fe^{3+}$ is around 0.7% smaller than that of the octahedron hosting the $Mg^{2+}$. The lattice expansion is attributed to the increase of the volume of the tetrahedron $T_{Si}$, which hosts the $Al^{3+}$ ion, added for charge compensation in place of Si. The volume of $T_{Si}$, $V(T_{Si})$, increases of 21% in comparison to the $V(T_{Si})$ in the pristine geometry, when $T_{Si}$ hosts the $Si^{4+}$ ion. When the $Fe^{3+}$ impurity is at the tetrahedron site tt, the volumes of the octahedra sites $Oc^1$ and $Oc^2$ differ in about 0.5%, and the volume expansion is attributed to the increase in the volume of the tetrahedron tt, V(tt), by about 30%. These results agree with the expected effective ionic radii differences of the ions[71]. These impurity-induced geometry modifications are expected to impact the electronic structure of phlogopite crystals. In particular, the volume changes of the octahedra and tetrahedral where the Fe defects take place are directly related to the changes in the bond lengths between Mg and O to Fe and O atoms (see more details in the Supporting Information).

We now analyze the bandgap of phlogopite crystals without and with Fe ions. Figure 4 presents the density of states (DOS) of phlogopite crystal considered in Figure 3. The electronic structure of pristine phlogopite is displayed in Figure 4a. Panels *I* and *IV* in Figure 4a show the total DOS of pristine phlogopite $Mg_6(Al_2Si_6)O_{20}(OH)_4K_2$. The valence band edge in pristine phlogopite is mainly composed by O *p*-states, while the conduction band edge is mainly composed by states from O (~34%), K (~30%), and Mg (~18%) (see Supporting Information for more details on the pristine phlogopite electronic structure). The bandgap of pristine phlogopite, calculated by using Scuseria-Ernzerhof hybrid functional (HSE) is approximately 7.0 eV (panel *I* in Figure 4a), in good agreement with experimental evidence in synthetic phlogopite[67]. The partial DOS of the Mg atom, the O atoms bound to H (from the OH) and to Mg atoms, {H}-O-{Mg}, and the O atoms bound to Mg atoms only {Mg}-O-{Mg} of the octahedron site $Oc^1$ are presented in the panels *II-IV*, respectively (see the labels above the graphics). The contribution of the *s* and *p* orbitals are also indicated in these graphics, by red and green lines, respectively, and their sum in black lines. Panels *VI-VIII* in Figure 4a reveal the analyzes for the octahedron $Oc^2$ site.

Panels *I* and *IV* in Figure 4b present the electronic structure of $Mg_5Fe(Al_2Si_6)O_{20}(OH)_4K_2$ resulting due to the isovalent substitution of $Mg^{2+}$ by $Fe^{2+}$ ion in the $Oc^1$ and $Oc^2$ sites, respectively. The $Fe^{2+}$ ion induces similar electronic structure modification, regardless in the $Oc^1$ (panels *II-IV*) or $Oc^2$ (panels *VI-VIII*) sites, introducing *d*-Fe (*d*-states are indicated by blue lines in Figure 4b) and *p*-O states, within the pristine phlogopite bandgap, in the same energy range, due to the Fe-O bonding formation. As a consequence, the phlogopite bandgap is strongly reduced, becoming equal to 3.55 eV. This value is in agreement with our experimental observation (inset in Figure 2c). Panels *I* and *IV* in Figure 4c represent the electronic structure of $Mg_5Fe(Al_3Si_5)O_{20}(OH)_4K_2$ resulted from the substitution of $Mg^{2+}$ by $Fe^{3+}$ ion in the $Oc^1$ and $Oc^2$ sites, respectively. In these cases, the hybridization of the *d*-Fe and *p*-O orbitals occurs mainly near the conduction band edge. The reduction of the phlogopite bandgap is smaller than in the case of $Fe^{2+}$ ions at the octahedra, becoming equal to 5.85 eV.

Panel *I* in Figure 4d displays the DOS of pristine phlogopite, whereas panels *II-IV* show the partial DOS of the Al atom, the O atoms bound to Si and the Al atoms, {Si}-O-{Al}, and the O atoms bound to Mg and the Al atoms only {Mg}-O-{Al} of the tetrahedron site tt, respectively (see the labels above the graphics). Panels *V-VIII* in Figure 4d reveal the details of the electronic structure of the corresponding sites in the $Mg_6(AlFeSi_6)O_{20}(OH)_4K_2$ geometry, resulting due to the isovalent substitution of $Al^{3+}$ by $Fe^{3+}$ ion in phlogopite, at the tetrahedron site tt. The presence of $Fe^{3+}$ ions in the tetrahedral site leads to bonding between *d*-Fe and *p*-O orbitals, introducing states mainly near to the conduction band edge. The modifications in the electronic structure of phlogopite resemble that observed in the case of $Fe^{3+}$ in the octahedral sites $Oc^1$ and $Oc^2$. In the latter cases ($Fe^{3+}$ in $Oc^1$ and $Oc^2$ sites), two new narrow bands formed by *d*-Fe and *p*-O states are near the conduction band edge. While in the former ($Fe^{3+}$ in tt site), the corresponding new band closest to the conduction band edge is wider in energy (see the Supporting Information for more details on the electronic structure of these bulk compounds, as bandgap per spin channel, and partial DOS for all atoms). Therefore, our results reveal that the substitution of $Mg^{2+}$ by $Fe^{2+}$ ions is the main responsible for the reduction of the effective optical bandgap in phlogopite. In the supplementary information, we show DFT results on 1L-phlogopite, demonstrating that its bandgap gets narrower with reducing the number of layers, from 7 eV in bulk, down to 6.2 eV in the 1L limit. This reduction is attributed to the interaction between the K and the O atoms from the tetrahedral layer, which becomes stronger in comparison to the interaction in the bulk form. In addition, the effects of $Fe^{2+}$ ions in the $Oc^1$ and $Oc^2$ sites in to the 1L-phlogopite electronic structure and geometry are also investigated, showing similar trends to the bulk results.

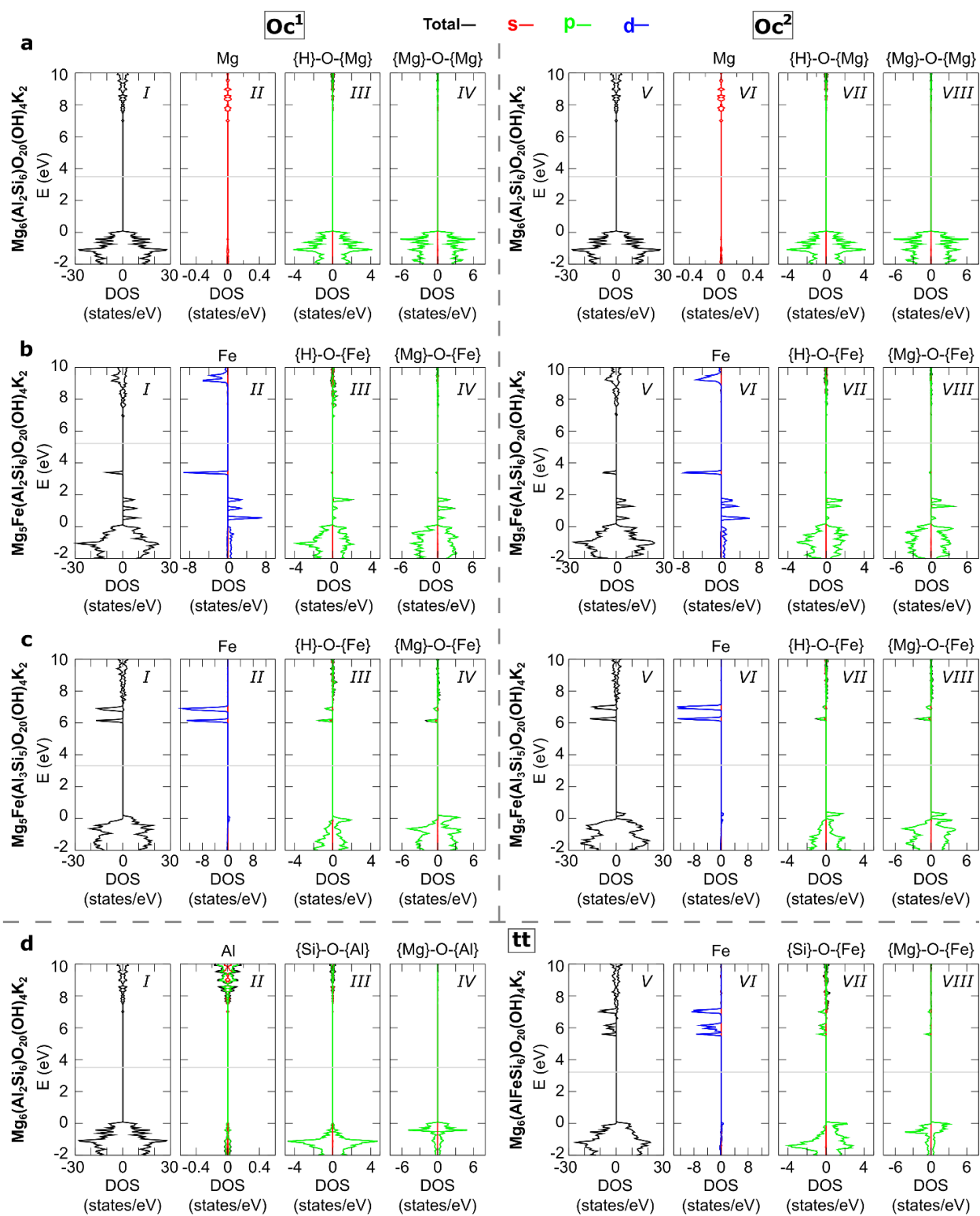

**Figure 4:** First-principles calculation of bulk Fe:phlogopite structures. Electronic structure of (a) pristine, (b) Oc$^1$:Fe$^{2+}$ and Oc$^2$:Fe$^{2+}$, (c) Oc$^1$:Fe$^{3+}$ and Oc$^2$:Fe$^{3+}$, and (d) pristine and tt:Fe$^{3+}$ phlogopite crystals. The graphics in panels *I* and *V* present the total DOS of phlogopite crystals. Contribution of the atoms in the octahedral sites Oc$^1$ and Oc$^2$,

Mg, Fe and O, and of atoms in the tetrahedron site tt, Al, Fe and O are indicated in the adjacent panels, where red, green and blue lines indicate the orbital contribution of s, p and d-states, and their sum are indicated by black lines, respectively. States below (above) the horizontal gray line are occupied (empty) and the up (down) spin-polarized states are represented as positive (negative) DOS. The calculations were performed by using HSE hybrid exchange-correlation functional.

In the following, we use the fact that phlogopite has weak van der Waals forces between the layers and strong in-plane bonds[21] to apply the mechanical exfoliation technique[1] for obtaining ultrathin slabs. Figure 5a (5b) shows the optical image of several phlogopite (hBN) flakes obtained by direct exfoliation of a bulk-crystal onto $SiO_2/Si$. The apparent color of the flake under the optical microscope depends on its thickness[72]. The lateral flake-size depends directly on the exfoliation process (scotch tape repetition), then, it may vary from sample to sample. However, it is possible to identify flake sizes varying from hundreds of nanometers up to hundreds of microns (see Figure S5). Moreover, the flake-thickness follows the same behavior, but one can easily identify flakes ranging from about 1 nm thick (thickness of an expected[25,30] 1L) up to hundreds of nanometers. It is known that LMs are able to conform to the topography of the underlying substrate[73,74], hence, we perform topographical analysis of the exfoliated flakes using AFM (see Methods for details). We identify that, as for hBN[74], phlogopite shows atomically flat (root mean square (rms) roughness, $R_{rms} < 0.2$ nm) surfaces over large areas. Figure 5c-d shows AFM topography images measured on phlogopite (58 nm thick) and hBN (54 nm thick) flakes, respectively. The average roughness measured in our samples are $R_{rms} = (0.17 \pm 0.03)$ nm (phlogopite) and $R_{rms} = (0.12 \pm 0.02)$ nm (hBN). These results indicate that phlogopite crystals show good surface topography and could potentially be embedded in vdWHs. Furthermore, the large area of these nanometer-thick phlogopite sheets, comparable to one of the exfoliated hBN slabs, makes it promising candidates for use not only as substrates in LM-based devices but also as insulator barriers in planar tunnel junctions or as dielectrics in field-effect transistors and nanocapacitors. Also, phlogopite presents a DC dielectric constant similar to hBN, as well as a lower refractive index with in-plane and out-of-plane anisotropies, all of which can be exploited for exciton- and plasmon-based vdWHs in nanophotonic and optoelectronic devices.

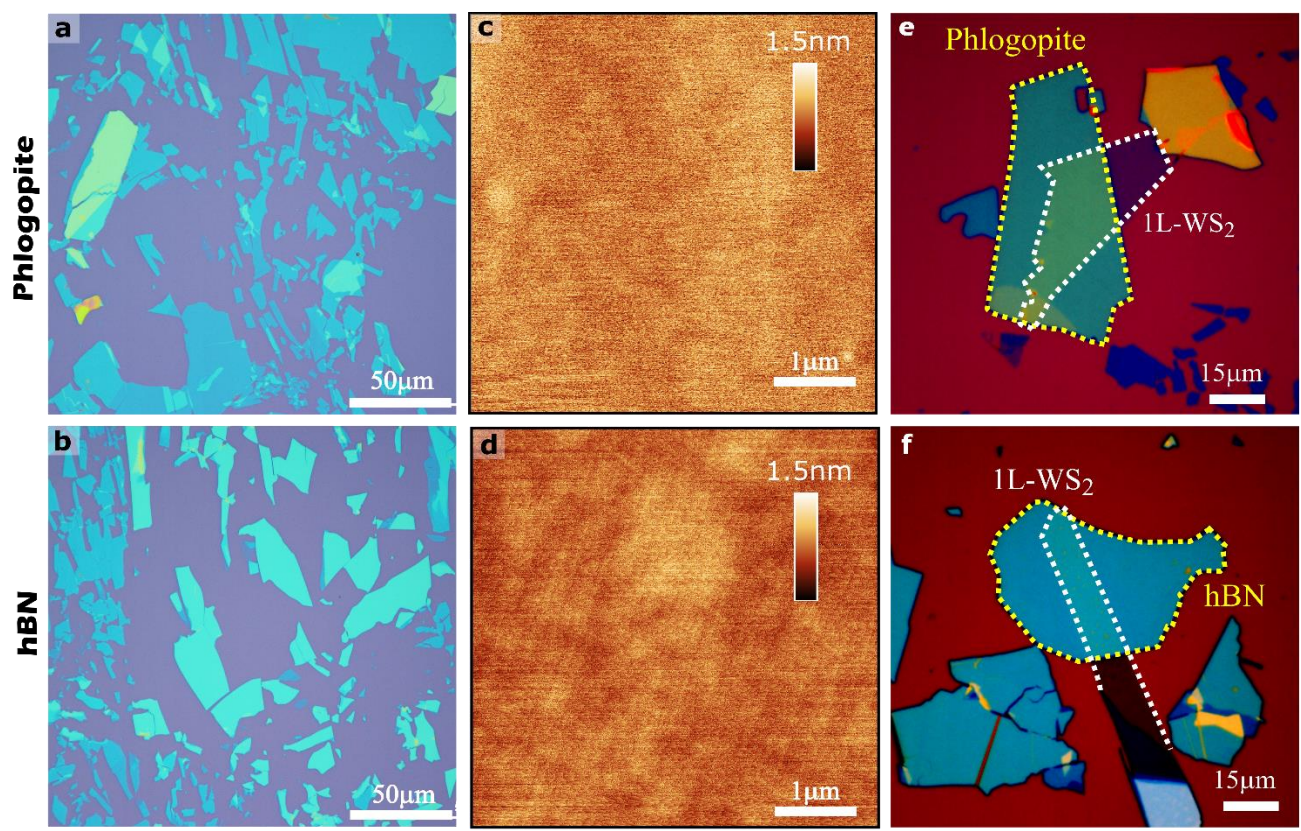

**Figure 5:** Phogopite flakes in van der Waals heterostructures. Optical micrograph of several (a) phlogopite and (b) hBN flakes exfoliated onto 285nm $SiO_2$/Si substrate. The different colors of the flakes correspond to different thicknesses. AFM topography images of ultrathin (c) phlogopite and (d) hBN slabs. Optical micrograph (false color) of a 1L-$WS_2$ transferred onto selected (e) phlogopite and (f) hBN flakes. In Figures (e) and (f) the dashed lines mark the LMs edges.

After the mechanical exfoliation, we have also investigated the stability of phlogopite under thermal annealing in air and in $H_2$/Ar atmosphere (see Methods for details). The stability of the LMs in air is a crucial processing issue since it allows the manipulation of the materials without the need of sealed glovebox systems, making it easier to handle[75]. Moreover, understanding if the LMs are stable under a certain temperature is also crucial for vdWHs fabrication because the LMs are often annealed either in air or gas atmosphere for both cleaning the material surface and ensure the best layer-to-layer and layer-to-substrate adhesion, eliminating a substantial portion of air pockets (bubbles) on the interfaces between the constituent layers. Figures S5b and S5c present the optical images of the flakes submitted to thermal annealing in oxidizing and reducing atmospheres, respectively. The results demonstrate that phlogopite flakes are well-stable at least up to 300 °C (maximum temperature analyzed in this work) and show no signs of topography alteration. AFM measurements were performed in annealed and non-annealed flakes and no significant changes in roughness were observed (data not shown). It is also important to stress that the flakes were left in ambient conditions for at least 9 months and exhibited no signs of degradation (Figure S5a). Therefore, our results suggest that phlogopite is strongly stable under ambient environment and thermal process, which is a good requirement for LMs-based applications.

We now focus on the use of this material as atomically flat substrates in vdWHs. Several works have shown that the underlying substrate affects the excitonic emission in monolayer transition metal dichalcogenides (1L-TMDs), thus changing the optical performance of the devices designed[76–79]. Figure 5e-f shows optical images of the transferred 1L-WS$_2$/phlogopite and 1L-WS$_2$/hBN materials, respectively (see Methods for description details regarding the vdWHs fabrication). For clarity, the optical images have false color contrast, hence, one can distinguish the 1L-WS$_2$ flakes from the underlying LMs. The 1L-WS$_2$ flakes are intentionally deposited partially over the LMs and SiO$_2$, thus, we can directly compare the effect of the underlying LM to the SiO$_2$/Si region. Thereafter, PL spectroscopy is applied to investigate the effect of the substrate on the PL emission of 1L-WS$_2$ (see Methods for details). Figure 6a plots the representative PL spectra for 1L-WS$_2$ at room temperature on the different substrates. The normalized spectra show the characteristic features from 1L-WS$_2$ close to 617 nm (2.01 eV) and 630 nm (1.97 eV). These signatures are commonly associated with the emission from neutral exciton (X$^0$) and negatively charged trions (X$^-$) because both exciton species can coexist at room temperature[76–79]. The findings in Figure 6a indicate that the underlying substrate affects the X$^-$ emission. By fitting the data with Lorentzian functions (not shown), we extract quantitative information about the X$^0$ and X$^-$ emissions, as shown in Figure 6b-c. It is worth noticing that changes in position and spectral weight (X$^0$/X$^-$ integrated area ratio) of the peaks in the PL spectrum are affected by external doping[80]. On the one hand, no significant variations are observed for the position of both peaks, suggesting no changes in the bandgap and exciton binding energy of 1L-WS$_2$ by the LM-substrate[81]. The small shift of the PL peak position from sample to sample could be related to some strain induced by the PDMS transfer. Conversely, Figure 6c shows a clear difference in the relative spectral integrated area ratio of exciton and trion (X$^0$/X$^-$). Normally, the formation of X$^-$ suppresses the radiative recombination of X$^0$ and thus reduces the X$^0$/X$^-$ PL ratio[80,81]. Therefore, our results suggest that the SiO$_2$ substrate induces extra n-type doping to the 1L-WS$_2$ flakes, consequently, increasing the X$^-$ formation. Whereas phlogopite flakes work similarly to hBN crystals, decreasing the contribution from the $SiO_2$, and favoring the radiative recombination ($X^0$) most likely due to suppression of charge transfer from the $SiO_2$ surface[77,78]. Our Raman data (not shown) performed on these samples show negligible changes to the Raman modes, suggesting no detectable induction of extra doping coming from the phlogopite substrate. Note that all flakes are selected by optical contrast followed by AFM characterization, and the thickness of all selected LM-substrate is in the range of 30-60 nm. This thickness selection is needed to decrease the effect of the charged puddles (trapped impurities) at the $SiO_2$ surface that is known to affect the electronic properties of the LMs[76–79]. Consequently, our results suggest that such material thickness is enough to improve the $X^0$ recombination efficiency of 1L-TMDs. Although low-temperature measurements are not performed and it is out-of-scope of our work, further experiments (electrical and optical) using phlogopite as dielectric substrate in low-temperature gated-PL (field-effect transistor architecture) will be employed to understand the influence of the Fe impurities to the electronic properties of the transferred and active 1L-TMD. Finally, we must highlight that all features discussed in our work (large bandgap material, ultrathin layers, atomically flat surface, and air stability) would make phlogopite a promising candidate for using in ultrathin and flexible low-cost LM-based optoelectronic devices.

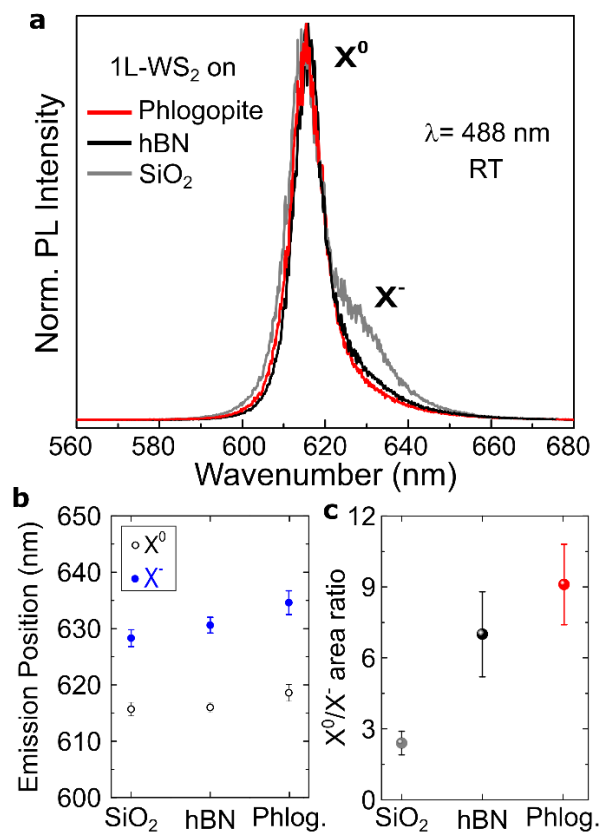

**Figure 6:** Photoluminescence characterization of 1L-WS$_2$/phlogopite heterostructures. a) Normalized PL spectra of 1L-WS$_2$ on phlogopite, hBN and SiO$_2$ substrates collected under 488nm excitation and laser power ~100μW at room temperature (RT). The neutral exciton (X$^0$) and trion (X$^-$) emission are indicated. b) Peak position of X$^0$ and X$^-$ emission for different substrates. c) X$^0$/X$^-$ area (integrated) ratio for different substrates.

**Conclusion**

We systematically studied the basic properties of ultrathin phlogopite crystals by means of XRD, XRF mapping, WDS, EPR, UV-Vis, AFM, FTIR, Mössbauer, Raman, and DFT calculations. We showed that mono- and few-layers of this material are air and temperature stable, are easily obtained by standard mechanical exfoliation technique, and have atomically flat surface. Moreover, our results confirmed that natural phlogopite presents Fe ions in its lattice and this explains the reduction observed for its optical bandgap (down to 3.6 eV). Nevertheless, the existence of external impurities could boost its use on defect engineering for localized luminescence experiments. We then fabricated ultrathin 1L-WS$_2$/phlogopite heterostuctures and demonstrated that the X$^0$ recombination efficiency is at least 3-fold higher than on $SiO_2$ substrates, similarly to those obtained on $1L-WS_2$/hBN heterostructures. Finally, the present proof-of-concept study shows that phlogopite can be easly exfoliated down to mono- and few-layers with large areas (thousands of $\mu m^2$), concluding that this mineral should be regarded as good and promising candidate for LM-based applications as a low-cost layered nanomaterial.

### Methods

**X-ray Diffraction:** The experiments were acquired in a commercial diffractometer (XRD Rigaku, Miniflex II) with Cu-K$_\alpha$ radiation ($\lambda$ = 1.5418 Å), using a 4-crystal monochromator and collimating slits in the Bragg-Brentano geometry ($\theta$:2$\theta$). The generator voltage and current were set to with 30 kV and 15 mA, respectively. For acquiring the diffractogram a powder sample was made from the mechanical grind of a large crystal (inset in Figure 1a).

**Wavelength-dispersive spectroscopy:** For the WDS measurements, a phlogopite sample was prepared on a resin support, polished and metalized with carbon. The quantitative microanalysis by WDS was performed for several points at different regions of the sample for a statistical quantification using a previously calibrated Jeol 8900 electron microprobe.

**Synchrotron radiation X-ray fluorescence:** Synchrotron radiation XRF mapping in an area of 250 x 250 $\mu m^2$ and 5 x 5 $\mu m^2$ was performed at room temperature with beam size down to 500 x 200 $nm^2$ using SDD 4 elements Vortex (Hitachi) detectors and excitation energy of 7.2 keV selected as the spectral region of interest for Fe K-alpha emission line with an energy resolution of $\Delta E/E = 10^{-4}$ at the Tarumã endstation / Carnauba beamline from the Sirius Synchrotron source. The XRF mapping images were analyzed by the PyMCA software[82]. The brighter red signal indicates the higher Fe concentration areas in phlogopite.

**Electron Paramagnetic Resonance:** For the EPR measurements, a custom build commercial Magnettech MiniScope MS 400 X-band spectrometer coupled to water-cooled electromagnets capable of producing magnetic fields up to 800 mT was used. The EPR angular dependence at room temperature of a bulk phlogopite sample coupled to a goniometer was performed with *c* axis perpendicular to the rotating axis, in which the *c* axis will be parallel to the applied magnetic field for a specific orientation. The parameters used to acquire all spectra were 9.440(1) GHz with approximately 30 mW of power, 100 kHz of modulation, and 0.5 mT of field modulation. The EPR signals are labeled by their effective g values defined by the relation $hv = g\beta B$, where h is Planck's constant, $v$ is the microwave frequency, $\beta$ is the Bohr magneton and $B$ is the modulus of the magnetic field.

**Optical absorption:** The absorbance spectrum and the reflectance spectrum needed to produce the Tauc plot were acquired with a Shimadzu 3600 Plus UV-Vis-NIR spectrometer on a phlogopite bulk with dimensions (~0.1 x ~5 x ~10 mm$^3$ – thickness x width x length) fixed to the sample holder. This sample was carefully peeled off with tweezers from the crystal shown in Figure 1. The grating change occurs at 720 nm for the spectral range from 250 to 1250 nm during the experiment.

**Mössbauer spectroscopy:** The Mössbauer spectrum was measured at 300 K in constant acceleration mode (triangular velocity) using a $^{57}$Co source in Rh matrix. Phlogopite powder was pressed into a sample holder (Ø=0.9 cm) to produce an area of ~0.6 cm$^2$. Sample thickness was close to the thin absorber approximation, to avoid thickness effects. The spectrum was fitted with Lorentzian doublets using Normos software[83]. The full width at half maximum (FWHM) considered for all doublets was 0.35 mm/s. The isomer shifts and quadrupole splittings were allowed to vary to adjust correctly the spectrum.

**Raman and Photoluminescence spectroscopy:** The experiments were done in a confocal Raman microscope WITec Alpha 300R in the backscattering configuration. We used 488nm excitation laser with a 100x objective which has a numerical aperture (NA) of 0.9. For the phlogopite Raman spectra, the laser power was kept constant at 10 mW during all measurements. Each Raman spectrum was collected with five accumulations of 60 sec with a grating of 1,800 grooves/mm. For the PL measurements on the 1L-WS$_2$ flakes transferred to hBN and phlogopite, we used 100 µW to avoid damaging of the 1L-WS$_2$. Each PL spectrum was collected with three accumulations of 10 sec with a grating of 600 grooves/mm. Several PL spectra were collected on the same sample (*i.e.,* 1L-WS$_2$ partially transferred onto the LMs) to check uniformity and different LMHs of each insulator were used to estimate the impact of the underlying substrate to the PL emission. The data presented in Figure 6b-c were averaged considering 15 PL spectra of each material. Previously to all measurements, the 488 nm laser was aligned on a prime silicon wafer, using the Si peak at 521 cm$^{-1}$ as reference.

**Fourier-transform infrared spectroscopy:** The micro-FTIR measurements were carried out in a Cary 620 FTIR microscope from Agilent Technologies using a Globar source with 8 cm$^{-1}$ of spectral resolution. Due to the non-transparency of the sample, single-point spectra in reflectance mode of a crystal fragment from the bulk phlogopite sample placed atop a flat gold surface were recorded by a MCT detector and a 25x objective lens with approximately 400 x 400 $\mu m^2$ of spot size. The final spectrum is an accumulation of 64 scans normalized by the spectrum of a clean Au surface as background reference.

**Exfoliation and assembly of the van der Waals heterostructures:** For the assembly of the LMHs we have used the following procedure: i) phlogopite and hBN bulk-crystals were mechanically cleaved[1] individually by scotch tape onto 285 nm $SiO_2$/Si substrates. The crystals were checked under the optical microscope and the selected thickness of all crystals was chosen based on optical contrast. ii) $WS_2$-bulk crystal (sourced from HQ Graphene) was exfoliated by scotch tape, then, exfoliated again on a polydimethylsiloxane (PDMS) stamp placed on a glass slide for inspection under an optical microscope. Optical contrast was used to identify 1L-$WS_2$ prior to transfer[84]. iii) 1L-$WS_2$ flakes onto the PDMS stamp were then aligned on the selected phlogopite (or hBN) flakes and stamped using xyz micromanipulators[85].

**Atomic Force Microscopy:** The AFM experiments were obtained with a Bruker Dimension Icon system operated in tapping mode. To characterize all samples, the thickness and roughness of the crystal's slabs were estimated from the topography images collected in 5 x 5 $\mu m^2$ scan area and with 512 scanlines under ambient conditions. The $R_{rms}$ values presented in this work were averaged based on at least five flakes of each material with a minimum flake thickness of 20 nm, decreasing, therefore, the roughness contribution coming from the $SiO_2$/Si substrates ($R_{rms}$= (1.3 ± 0.3) nm).

**Thermal annealing:** For the thermal annealing experiments, two approaches were pursued: i) in air – the $SiO_2$/Si substrates with exfoliated crystals (hBN and phlogopite) were left on a standard hot-plate under ambient conditions for 1 h at 175 °C, followed by another 1 h at 300 °C. ii) in gas environment – distinct $SiO_2$/Si substrates with exfoliated crystals (hBN and phlogopite) were inserted into a quartz tube in a tubular furnace at 25 °C. Thus, we set a gas flow of 90 sccm (Argon) and 300 sccm ($H_2$), a final temperature of 300 °C, and a temperature rate of 15 °C/min. The samples were left at 300 °C for at least 4 h before turning the heater off. Hence, the samples were cooled down to 100 °C with continuous flow before opening the furnace lead and closing all gases off.

**Ab initio simulations:** First-principles simulations were performed by using spin-polarized density functional theory (DFT)[86,87] as implemented in the Quantum Espresso Distribution[88,89]. The valence electron-ion interactions were considered by using norm-conserving pseudopotentials[90] (pseudopotentials files H.pz-

mt_fhi.UPF, K.pz-mt_fhi.UPF, Mg.pz-mt_fhi.UPF, Fe.pz-mt_fhi.UPF, Al.pz-mt_fhi.UPF, Si.pz-mt_fhi.UPF, and O.pz-mt_fhi.UPF, available at Reference[91]. The exchange-correlation energy is evaluated within the Perdew-Zunger's local density approximation (LDA) for geometry optimization calculations and within Scuseria-Ernzerhof hybrid functional (HSE)[92–95] for band structure calculations. 1L-Phlogopite calculations were carried by using proper 2D open boundary conditions[96], with images separated by approximately 24 Å of vaccum. Forces on the atoms and stress on the lattice are lower than 0.1 mRy/bohr and 10 MPa, respectively. The plane-wave kinetic energy cutoff to describe the electronic wave functions (charge density) was set to 42 Ha (168 Ha). The brillouin zone integrations were performed within the Monkhorst-Pack scheme[97], by using 4x2x2 (4x2x1) Gamma centre k-point grids in the geometry optimization calculations for bulk (1L). The brillouin zone integrations to evaluate the exchange operator were performed on additional 4x4x2 (4x2x1) q-point grids[98–100] for bulk (1L) phlogopite.

## DATA AVAILABILITY

The datasets obtained and analyzed during the study are available from the corresponding authors on reasonable request.

## CODE AVAILABILITY

Scripts for the theoretical analyses are available from the corresponding authors on reasonable request.

## ACKNOWLEDGMENTS

All Brazilian authors thank the financial support from CAPES and CNPq. The authors are thankful to the Center of Microscopy at Federal University of Minas Gerais for providing the equipment and technical support for the WDS experiments and to the Carnauba and IMBUIA beamlines from LNLS/CNPEM for the XRF and FTIR studies, respectively. A.R.C. acknowledges the financial support through the Fundo Mackenzie de Pesquisa e Inovação (MackPesquisa No.: 221017) and the CNPq (No. 309920/2021-3). A.R.C., D.A.N., V.T.A. and C.J.S.de.M. acknowledge the FAPESP financial support (Grants No.: 2018/25339-4, 2018/07276-5, and 2020/04374-6). A.R.C., R.L.M.L., J.R.S., A.M., I.D.B. and C.J.S.de.M acknowledge the Brazilian Nanocarbon Institute of Science and Technology (INCT/Nanocarbono). R.O., A.M., R.M.P. and K.K. thank the FAPEMIG financial support. I.D.B., J.R.-S. and R.L.M.L. acknowledge the support from CNPq (Grants No.: 311327/2020-6, 312865/2020-1, 433027/2018-5, and 420364/2018-8). I.D.B. and J.R.-S. acknowledge the prize L'ORÉAL-UNESCO-ABC for Women in Science Prize – Brazil (2021 and 2017 editions, respectively). J.R.-S. and R.L.M.L. thank the support from FAPEMIG through the Grant No. APQ-01922-21, APQ-01980-18, RED-00185-16, and TEC-RED-00282-16. R.L.M.L. acknowledges the computational time at CENAPAD-SP, CENAPAD-RJ/LNCC, CENAPAD-UFC, SDumont supercomputer, DFI-UFLA, and LCC-UFLA. Growth of hBN crystals was supported by the Elemental Strategy Initiative conducted by the MEXT, Japan (Grants No. JPMXP0112101001), and JSPS KAKENHI (Grants No.: 19H05790, 20H00354 and 21H05233). All authors thank Professor Marco A. Fonseca from Federal University of Ouro Preto for supplying the phlogopite crystal, as well as Jessica Fonsaca and Caroline Brambilla de Aquino for providing initial assistance in the experiments at MackGraphe.

**COMPETING INTERESTS**

The authors declare no competing interests.

**ADDITIONAL INFORMATION**

Supplementary information. The online version contains supplementary material available at

This is the authors' version (pre peer-review) of the manuscript: A.R.Cadore et. al., 2D Materials 2022.

This has been published online at https://doi.org/10.1088/2053-1583/ac6cf4

# Supplementary Information

# Exploring the structural and optoelectronic properties of natural insulating phlogopite in van der Waals heterostructures

Alisson R. Cadore[1]*, Raphaela de Oliveira[2], Raphael L. M. Lobato[3], Verônica de C. Teixeira[4], Danilo A. Nagaoka[1,5], Vinicius T. Alvarenga,[1,5] Jenaina Ribeiro-Soares[3], Kenji Watanabe[6], Takashi Taniguchi[7], Roberto M. Paniago[2], Angelo Malachias[2], Klaus Krambrock[2], Ingrid D. Barcelos[4], Christiano J. S. de Matos[1,5]

[1] *School of Engineering, Mackenzie Presbyterian University, São Paulo, Brazil*

[2] *Physics Department, Federal University of Minas Gerais, Minas Gerais, Brazil*

[3] *Physics Department, Institute of Natural Science, Federal University of Lavras, Minas Gerais, Brazil*

[4] *Brazilian Synchrotron Light Laboratory (LNLS), Brazilian Center for Research in Energy and Materials (CNPEM), São Paulo, Brazil*

[5] *MackGraphe, Mackenzie Presbyterian University, São Paulo, Brazil*

[6] *Research Center for Functional Materials, National Institute for Materials Science, Tsukuba, Japan*

[7] *International Center for Materials Nanoarchitectonics, National Institute for Materials Science, Tsukuba, Japan*

*Corresponding author: alisson.cadore@mackenzie.br

## Section S1. Synchrotron radiation X-ray fluorescence in bulk phlogopite

Figure S1 demonstrates qualitatively that Fe ions are largely predominant compared with other trace elements.

The XRF spectrum is taken from a representative position in Figure 2a.

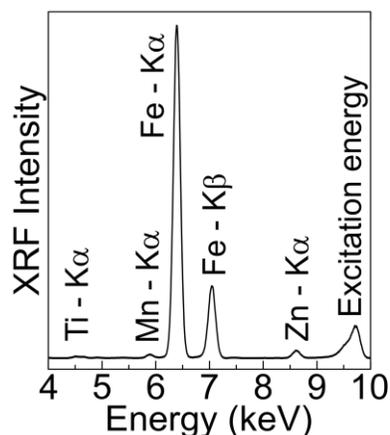

**Figure S1:** *a) XRF emission lines recorded with excitation energy of 9750 eV. Several trace impurities are identified but Fe is the major element.*

## Section S2. Raman and FTIR spectroscopy in bulk phlogopite

To further demonstrate the existence of Fe impurities in our phlogopite crystal, we perform Raman and FTIR measurements. Figure S2a shows a representative Raman spectrum of bulk-phlogopite. Our Raman spectrum displays 21 Raman modes located at approximately 103.8, 110.8, 116.1, 194.5, 229.0, 291.5, 332.6, 362.1, 381.1, 431.9, 451.6, 468.4, 508.0, 517.1, 676.2, 785.2, 791.8, 1014.9, 1050.5, 3663.4, and 3678.8 $cm^{-1}$. The spectrum collected agrees with references [1–4] and is close to the 27 Raman-active fundamental modes theoretically predicted [4]. Some spectral variations may be due to various reasons (*i.e.,* different spectrometers and laser excitation used to collect the data, distinct sample scattering conditions and resulting Raman tensor components that affect the spectra) but are mainly attributed to chemical and structural differences among the samples studied. Previous works [1–5] have debated that the position and shape of some Raman modes of phlogopite are strongly affected by changes in impurity concentration. For instance, the presence of Fe ions induce a peak splitting of the OH (3500-3700 $cm^{-1}$) and Si–O–Si (700-800 $cm^{-1}$) modes and the appearance of the strong peak in the 500-550 $cm^{-1}$ range [2,3]. More importantly, following the increase of Fe concentration in phyllosilicates, some Raman peaks exhibit a systematic position shift as shown in References [2,3]. Therefore, by comparing our data with references [1–4], we can affirm that our natural crystal presents Fe impurities. Figure S2b shows a representative IR spectrum of bulk-phlogopite. The FTIR absorbance spectrum shows two characteristic broad and strong peaks around 512 $cm^{-1}$ and 900 $cm^{-1}$, other less intense modes at about 655, 679, 732, and 804 $cm^{-1}$, as well as additional shoulders that are not clearly resolved. In phlogopite, all these modes are related to different vibrations (*i.e.,* bending or stretching modes of Si-O, Al-Si, Al-O, Al-O-Si, Mg-O, and others), hence, the existence of Fe (and other) impurities in its lattice is expected to modify the IR spectrum compared to pristine material. Note that the full description of IR modes can be found in detail in References [6–8], including a probable shift and splitting of the peak at 820 $cm^{-1}$ due to high $Fe^{2+}$ concentration and changes in peak intensity of the modes in the 600-850 $cm^{-1}$ range caused by other elements. Consequently, from previous work, it is clear that the IR spectrum of phlogopite can be different for samples obtained from diverse locations. Still, our IR spectrum is consistent with References [7,9] on natural crystals. Therefore, our experimental observations strongly support the existence of Fe impurities in the studied phlogopite material.

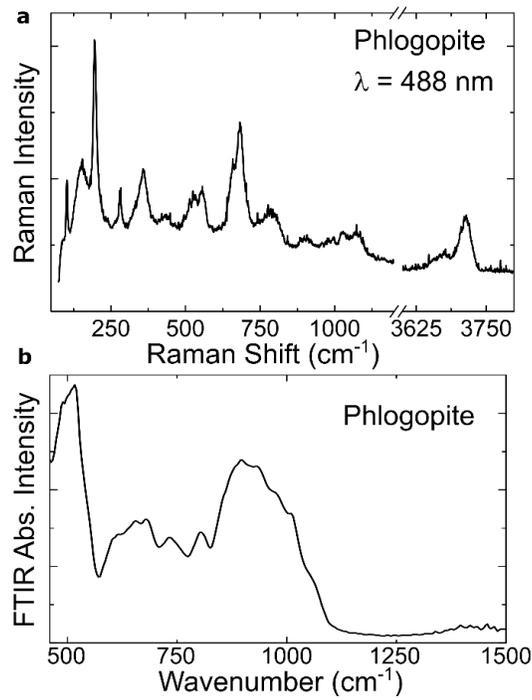

**Figure S2**: a) Representative Raman spectrum excited at 488 nm and b) FTIR absorbance spectrum of the bulk phlogopite sample.

## Section S2. Theoretical analysis of the geometries of pristine and Fe:phlogopite

Table 1 shows that the Fe defect induced lattice modifications in monolayer (1L) phlogopite are similar to that observed in bulk geometry, *e.g.,* the Fe ions cause expansion of the lattice vectors, and of the volume of the octahedron hosting the $Fe^{2+}$.

*Table 1: Monolayer (1L) phlogopite structural parameters for pristine (P), Fe:phlogopite at site $Oc^1$ ($Oc^1:Fe^{2+}$) and at site $Oc^2$ ($Oc^2:Fe^{2+}$). Lattice vectors (a,b) in angstrom, lattice angle ($\gamma$) in degrees and volume of the octahedral sites $Oc^1$ ($V(Oc^1)$), $Oc^2$ ($V(Oc^2)$), and tetrahedral sites $T_{Si}$ ($V(T_{Si})$) and $T_{Al}$ ($V(T_{Al})$), in angstrom$^3$, respectively.*

| 1L | a | b | $\gamma$ | $V(Oc^1)$ | $V(Oc^2)$ | $V(T_{Si})$ | $V(T_{Al})$ |
|---|---|---|---|---|---|---|---|
| P | 5.244 | 9.048 | 90 | 11.233 | 11.268 | 2.141 | 2.591 |
| $Oc^1:Fe^{2+}$ | 5.264 | 9.082 | 89.96 | 12.353 | 11.206 | 2.145 | 2.592 |
| $Oc^2:Fe^{2+}$ | 5.260 | 9.071 | 89.96 | 11.180 | 12.423 | 2.147 | 2.601 |

*Table 2: Average bond length between metal and oxygen atoms in the octahedral sites where Fe substitution take places in the $Oc^1:Fe^{2+}$, $Oc^2:Fe^{2+}$, $Oc^1:Fe^{3+}$ and $Oc^2:Fe^{3+}$ geometries in 1L- and bulk-phlogopite. We distinguished between the oxygen from the hydroxyl group M-O-{H} and the other oxygen atoms of the octahedron M-O-{Mg}. In the case of pristine (P) phlogopite the metal M is $Mg^{2+}$, and we took the average value considering both $Oc^1$ and $Oc^2$ sites. In the case of Fe:phlogopite, the metal M is $Fe^{2+}$ or $Fe^{3+}$, and we took the average value considering the Fe site.*

| | L1 | | | Bulk | | | | |
|---|---|---|---|---|---|---|---|---|
| | P | $Fe^{2+}$ | | P | $Fe^{2+}$ | | $Fe^{3+}$ | |
| | | $OC^1$ | $OC^2$ | | $Oc^1$ | $Oc^2$ | $Oc^1$ | $Oc^2$ |
| M-O-{H} | 2.039 | 2.115 | 2.110 | 2.010 | 2.085 | 2.089 | 2.034 | 2.046 |
| M-O-{Mg} | 2.049 | 2.115 | 2.121 | 2.040 | 2.112 | 2.115 | 2.032 | 2.025 |

In Table 2, we detail the bond length between the metal and the O atoms from the octahedron, belonging to the hydroxyl group (M-O-{H}), and to the O atoms shared by the adjacent octahedra (M-O-{M}) in both $Oc^1$ and $Oc^2$ octahedra sites. In pristine phlogopite, the metal M is $Mg^{2+}$, while in Fe:phlogopite, the metal M is $Fe^{2+}$ or $Fe^{3+}$. In the case of bulk phlogopite, the Mg-O-{H} is slightly smaller than the Mg-O-{Mg} bond. The presence of $Fe^{2+}$ impurities lead to increase of the M-O bonds by approximately 0.08A in the case of Fe-O-{H}, and by 0.07 in the case of Fe-O-{Mg}. Similar trends are observed in the case of 1L-phlogopite with Fe ions in the $Fe^{2+}$ valance states. For bulk-phlogopite with $Fe^{3+}$ ions, the Fe-O-{H} increases by 0.025 A, while the Fe-O-{Mg} decreases by 0.01A.

*Table 3: Average bond length between metal and O atoms in the tetrahedron site of bulk-phlogopite where Fe substitution takes place in the tt:$Fe^{3+}$ geometry. We distinguished between the oxygen bound to near silicon atoms M-O-{Si} and to magnesium atoms M-O-{Mg}. In the case of pristine phlogopite (P) the metal M is $Al^{3+}$. In the case of Fe:phlogopite, the metal M is $Fe^{3+}$.*

| | P | tt |
|---|---|---|
| M-O-{Si} | 1.738 | 2.867 |
| M-O-{Mg} | 1.6947 | 1.848 |

The isovalent substitution $Al^{3+}$ by $Fe^{3+}$ in the tetrahedron site of phlogopite leads to an increase of both the M-O-{Si} and M-O-{Mg} bonds, as shown in Table 3.

*Table 4: Average bond length between potassium and oxygen atoms from the tetrahedral in pristine and $Fe^{2+}$ impurities of the octahedral sites $Oc^1$ and $Oc^2$ in phlogopite. For pristine phlogopite, we also display the average difference between the z-coordinate of the K and O atoms. Lengths are given in angstrom.*

| | **L1** | | | **Bulk** | | |
|---|---|---|---|---|---|---|
| | P | $Fe^{2+}$ | | P | $Fe^{2+}$ | |
| | | $Oc^1$ | $Oc^2$ | | $Oc^1$ | $Oc^2$ |
| K-O | 2.558 | 2.577 | 2.573 | 2.685 | 2.700 | 2.702 |
| Dz (K-O) | 0.905 | | | 1.544 | | |

Finally, we analyze the influence of both dimensionality and iron $Fe^{2+}$ ions in the bond lengths between K and O atoms from the tetrahedral layer in Table 4. The dimensionality reduces the K-O bond length, in particular the component perpendicular to the tetrahedral layer. The lack of adjacent layers in the 1L, in contrast to the bulk geometry, results in stronger interaction between the K and O atoms from the tetrahedral layer. The average *z*-coordinate difference between the K and O atoms from the tetrahedral layers in the 1L is approximately 41.4 % smaller than that in the bulk. We attribute to this stronger K-O interaction the reduction of the pristine 1L-phlogopite bandgap with respect to the pristine bulk bandgap. The presence of $Fe^{2+}$ ions in the octahedral sites $Oc^1$ and $Oc^2$ have minor influence of the K-O bond length, which decreases by approximately 0.02 A in both 1L and bulk structures.

## Section S3. Theoretical analysis of the electronic structure of pristine and Fe:phlogopite

Figure S3 shows the electronic structure of pristine bulk and monolayer (1L) phlogopite, calculated by using LDA exchange-correlation functional. In the case of bulk phlogopite (Figures S3(a) and S3(b)), the valence band edge is mainly composed by oxygen *p*-states, while the conduction band edge is mainly composed by O, K, and Mg *s*-states. In the case of the 1L-phlogopite (Figures S3(c) and S3(d)), the lack of an adjacent layer allows a stronger interaction between the K and O atoms from the tetrahedral layer, in comparison to the bulk case. We attribute to the stronger K-O interaction the changes observed in the electronic structure of 1L-phlogopite, where the K s-orbitals increase their contribution to the conduction band edge, which shifts to lower energy region, reducing the bandgap from 4.83 eV (7.01 eV) in bulk to 4.75 eV (6.22 eV) in the 1L-phlogopite, as calculated by LDA (HSE) exchange-correlation functional.

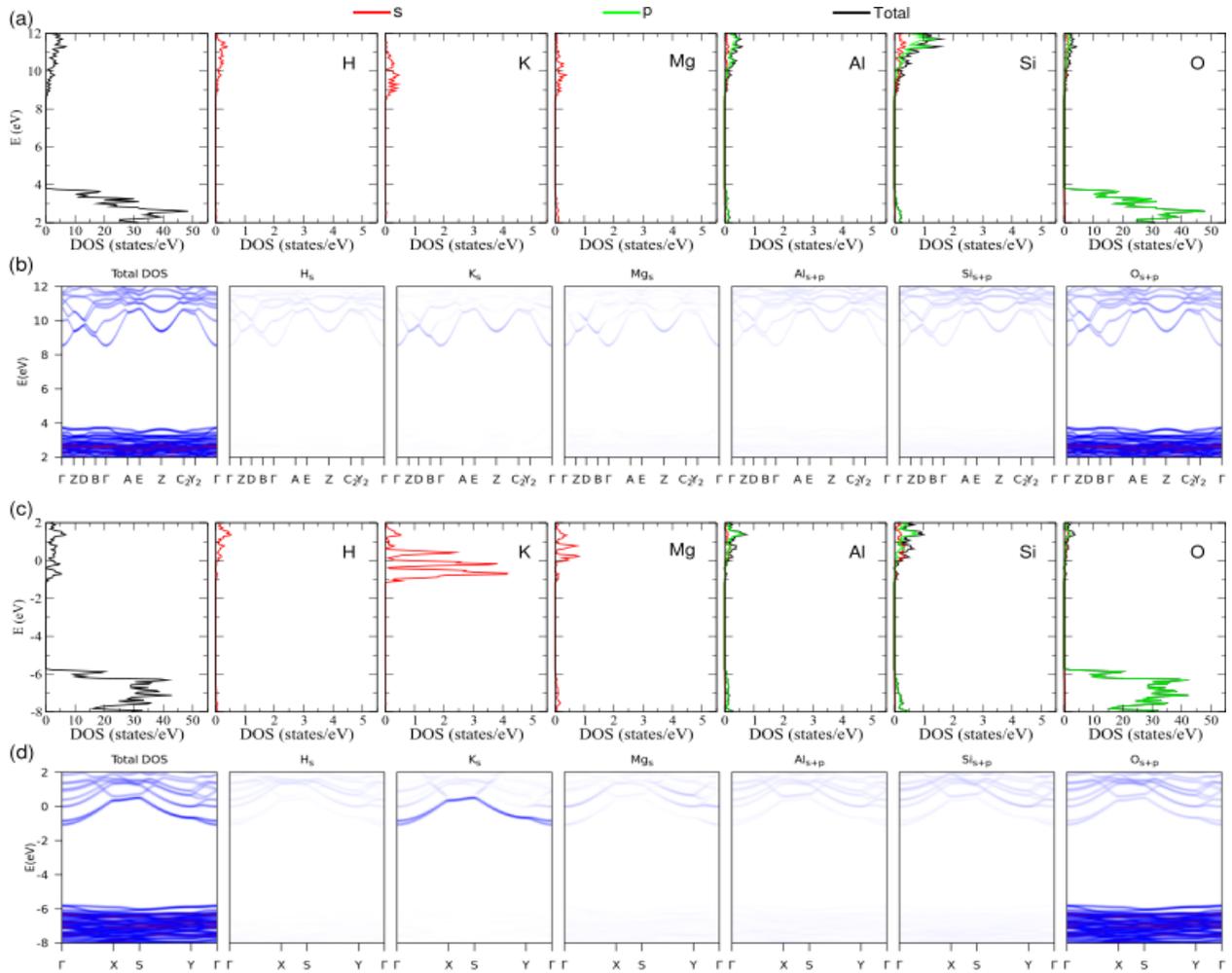

*Figure S3: Electronic structure of pristine bulk and 1L-phlogopite, as calculated by LDA. Bulk DOS (a) and projected DOS along high symmetry paths in the Brillouin zone (b). Monolayer DOS (c) and projected DOS along high symmetry paths in the Brillouin zone (d). The transition from null to maximum values of projected DOS follows the white-blue-red palette.*

Figure S4 displays the electronic structure of all considered bulk phlogopite and Fe:phlogopite compounds, as calculated by using HSE exchange-correlation functional.

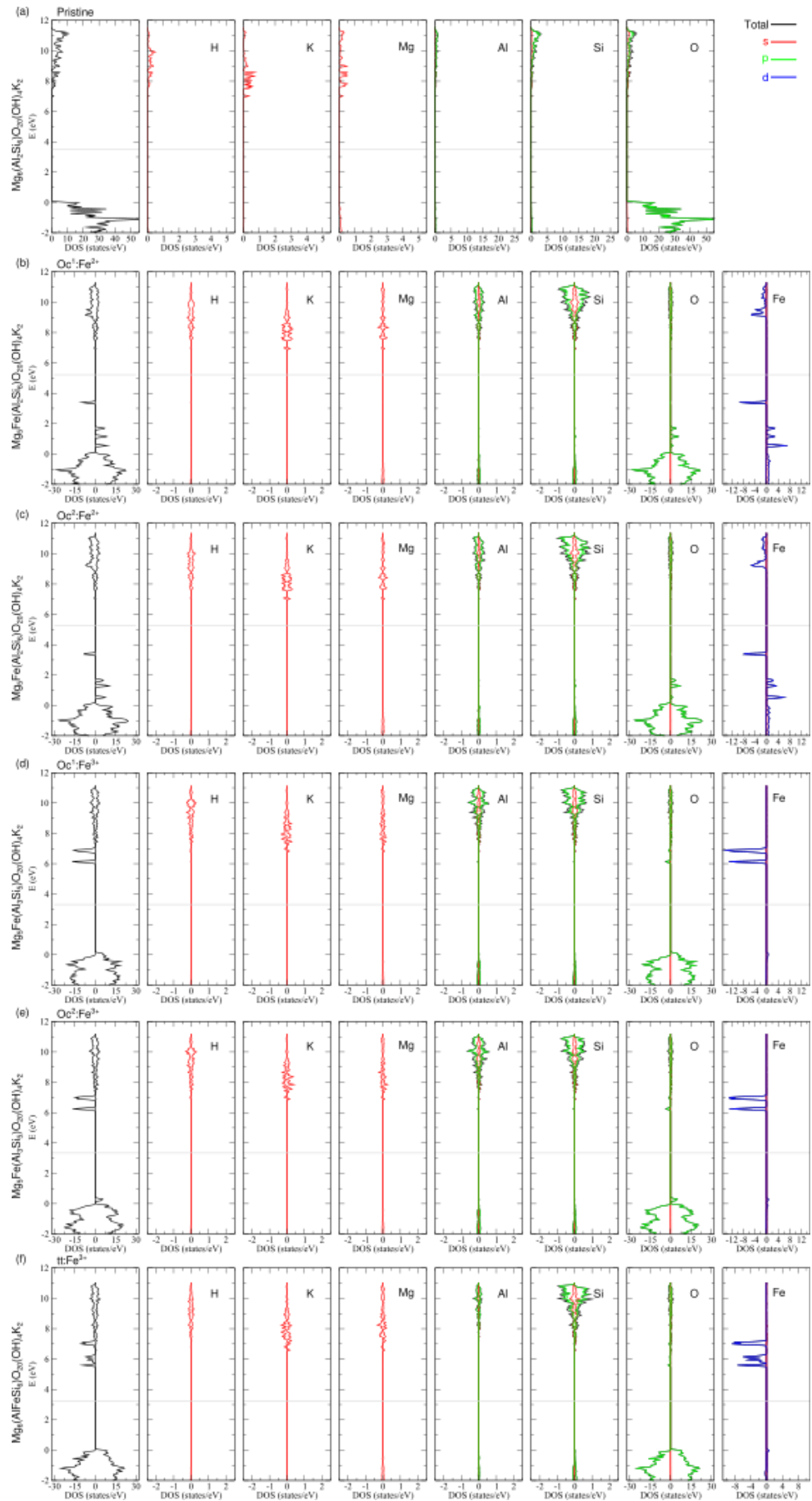

Figure S4: Electronic structure of bulk phlogopite and bulk Fe:phlogopite, as calculated by HSE exchange-correlation functional.

(a) Pristine, $Mg_6(Al_2Si_6)O_{20}(OH)_4K_2$.

(b) $Oc^1:Fe^{2+}$, $Mg_5Fe(Al_2Si_6)O_{20}(OH)_4K_2$.

(c) $Oc^2:Fe^{2+}$ $Mg_5Fe(Al_2Si_6)O_{20}(OH)_4K_2$.

(d) $Oc^1:Fe^{3+}$, $Mg_5Fe(Al_3Si_5)O_{20}(OH)_4K_2$.

(e) $Oc^2:Fe^{3+}$, $Mg_5Fe(Al_3Si_5)O_{20}(OH)_4K_2$.

(f) $tt:Fe^{3+}$, $Mg_6(AlFeSi_6)O_{20}(OH)_4K_2$.

Figure S5 displays the electronic structure of all considered 1L-phlogopite and 1L-Fe:phlogopite compounds, as calculated by using HSE exchange-correlation functional.

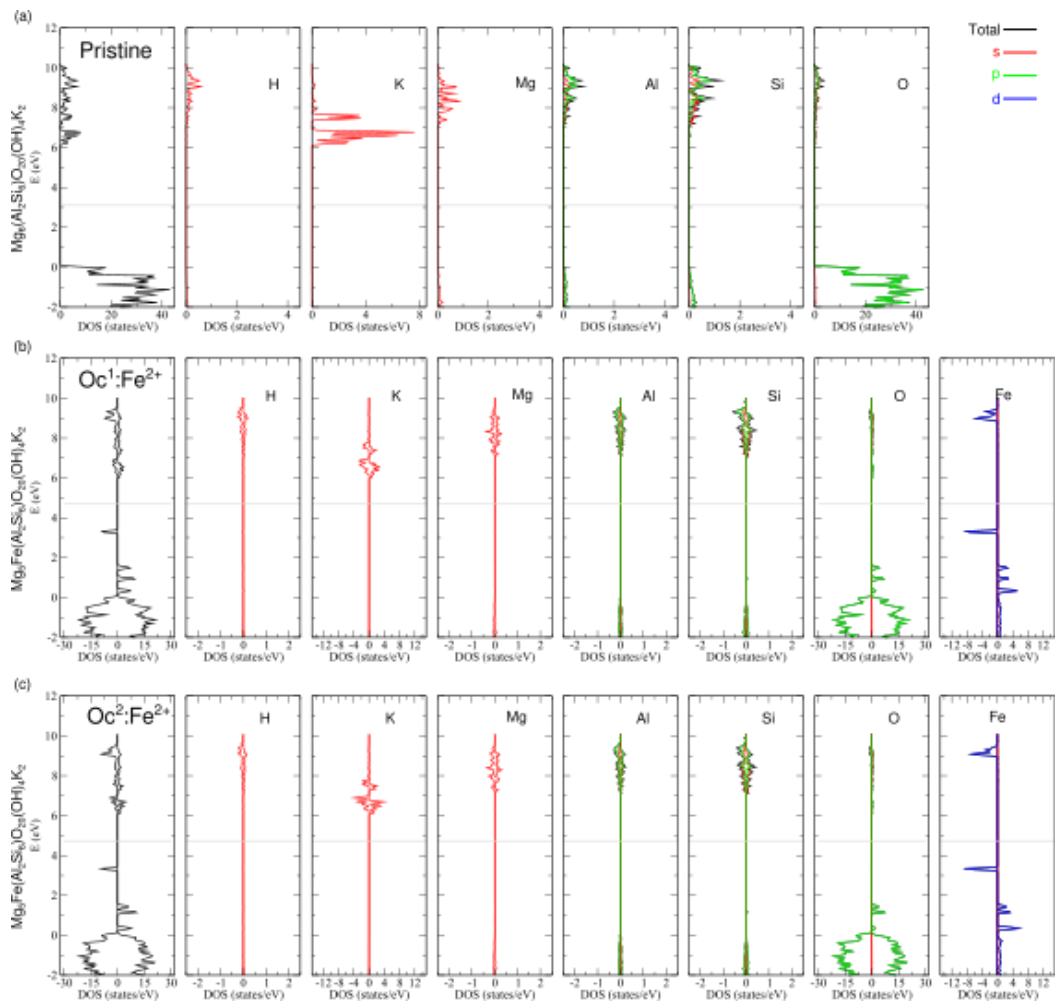

*Figure S5: Electronic structure of 1L-phlogopite as calculated by HSE exchange-correlation functional. (a) Pristine, $Mg_6(Al_2Si_4)O_{20}(OH)_4K_2$. (b) $Oc^1:Fe^{2+}$, $Mg_5Fe(Al_2Si_4)O_{20}(OH)_4K_2$. (c) $Oc_2:Fe^{2+}$, $Mg_5Fe(Al_2Si_4)O_{20}(OH)_4K_2$.*

Further details of the electronic structure in 1L- pristine and 1L-Fe:phlogopite are discussed next. Figure S6 shows in detail the influence of the Fe ions in the 1L-phlogopite electronic structure. The results are very similar to the calculated for bulk, where the Fe ions introduce Fe $d$-states within the bandgap of the pristine compound. We also show the contribution from K states, as the conduction band edge in the monolayers mainly composed by $s$-states from this atom. The results show that the Fe ions induce very small spin polarization in the s-band from K atoms.

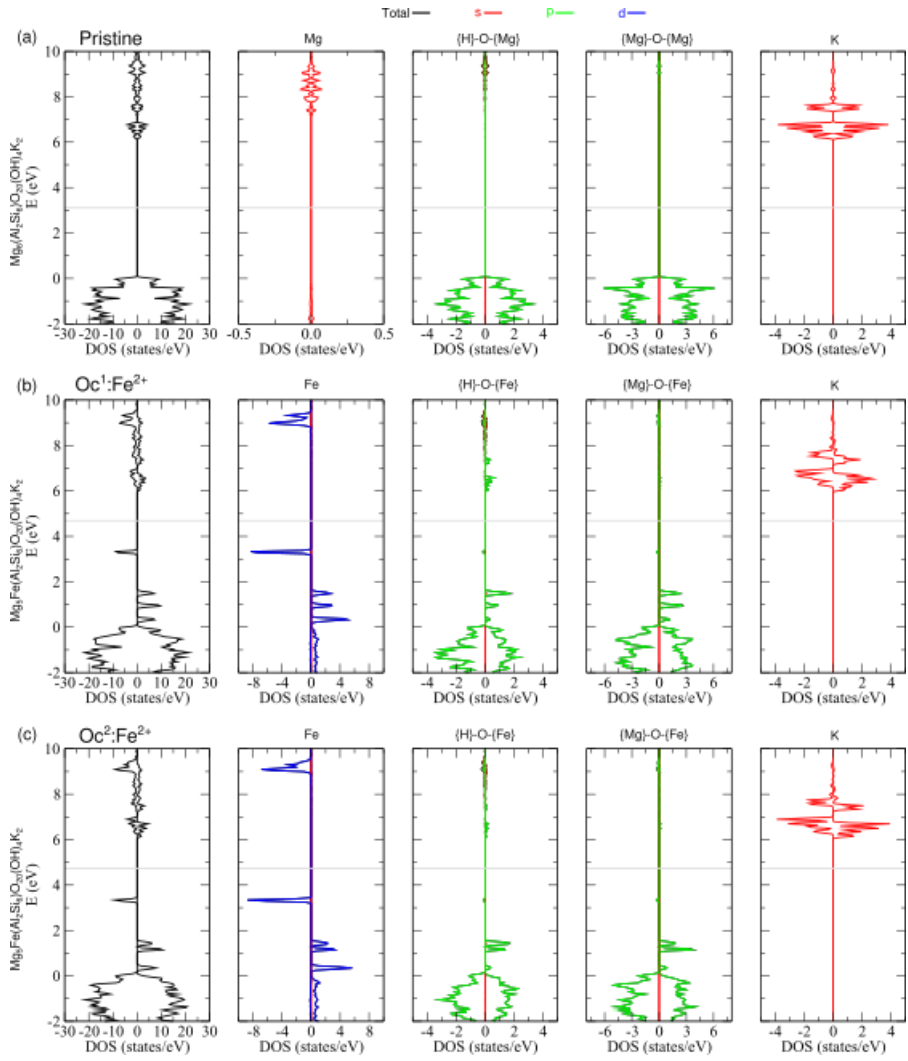

*Figure S6: 1L-phlogopite electronic structure. Contribution from the metal and O atoms in the octahedra sites hosting the Fe ions, or in the corresponding site in pristine structure (both $Oc^1$ and $Oc^2$ display the same electronic structure in pristine phlogopite). (a) Pristine. (b) $Oc^1$:$Fe^{2+}$. (c) $Oc^2$:$Fe^{2+}$. The calculations were performed by using HSE exchange-correlation functional.*

In Table 5, we show the details of the bandgap in the considered phlogopite compounds.

*Table 5: Bandgap of 1L and bulk phlogopite. We indicate values calculated by using LDA and HSE exchange-correlation functional, and per spin channel.*

| Bandgap | | 1L | | | Bulk | | | | | |
|---|---|---|---|---|---|---|---|---|---|---|
| | | P | Fe²⁺ | | P | Fe²⁺ | | Fe³⁺ | | |
| | | | Oc¹:Fe²⁺ | Oc²:Fe²⁺ | | Oc¹:Fe²⁺ | Oc²:Fe²⁺ | Oc¹:Fe³⁺ | Oc²:Fe³⁺ | tt:Fe³⁺ |
| LDA | Total | 4.75 | | | 4.83 | 0.75 | 0.79 | | | |
| | up | | | | | 3.18 | 3.30 | | | |
| | dw | | | | | 0.80 | 0.84 | | | |
| HSE | Total | 6.22 | 2.75 | 2.80 | 7.01 | 3.54 | 3.62 | 5.90 | 5.80 | 5.40 |
| | up | | 4.56 | 4.70 | | 5.26 | 5.37 | 6.65 | 6.50 | 6.40 |
| | dw | | 2.98 | 3.00 | | 3.58 | 3.67 | 6.15 | 6.20 | 5.50 |

**Section S4. Mechanical exfoliation and Environmental Stability of bulk-phlogopite**

Figure S7 shows representative optical images of phlogopite flakes obtained by mechanical exfoliation technique onto SiO$_2$/Si substrates. Figure S7a brings phlogopite flakes with several lateral-crystal size, ranging from hundreds of nanometers up to hundreds of microns. The flake-thickness follows the same behavior, and one can identify flakes ranging the 1 nm thick up to hundreds (bright orange/pink) of nanometers. Figures S7b and S7c present optical images of representative phlogopite flakes submitted to thermal annealing in air (1h @ 300°C) and in Ar/H$_2$ gas (4h @ 300°C), respectively. No signs of degradation are observed.

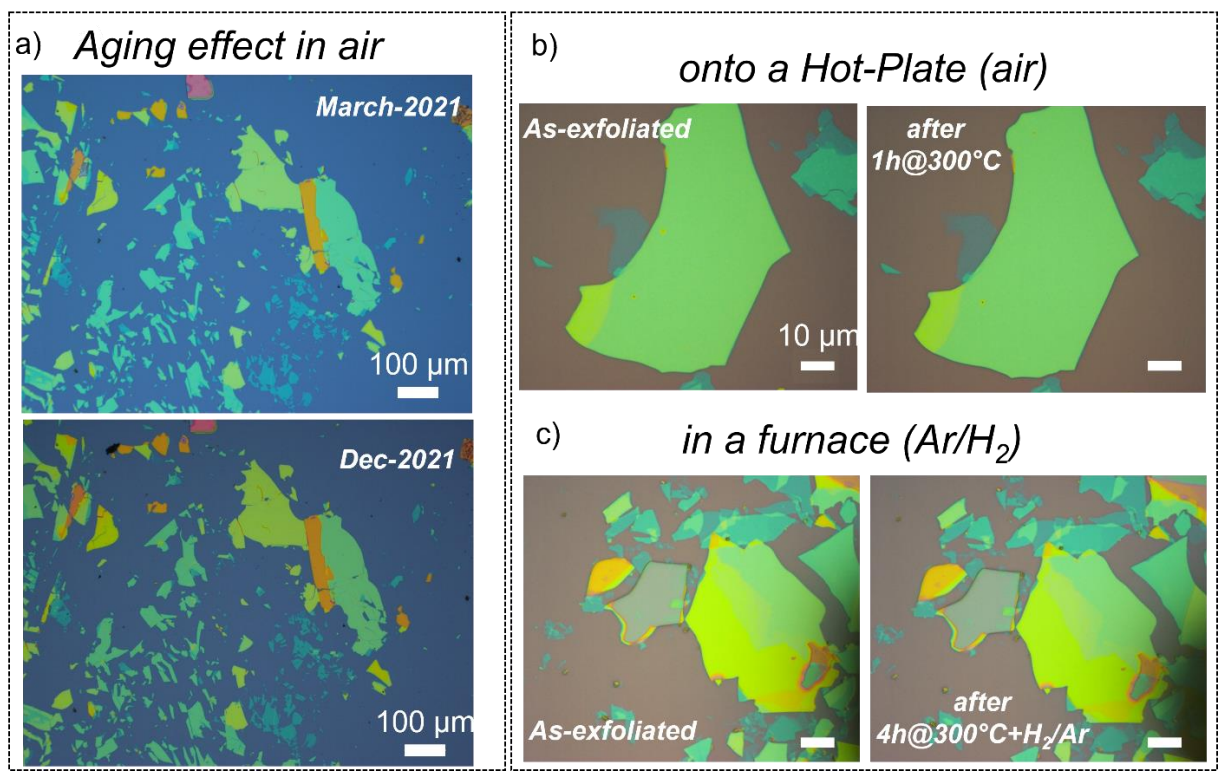

*Figure S7: a) Optical micrograph of several phlogopite flakes after mechanical exfoliation (top panel) and after nine months in air (bottom panel). b) Optical micrograph of phlogopite flakes after mechanical exfoliation (left panel) and after thermal annealing in air onto a hot plate at 300°C (right panel). c) Optical micrograph of phlogopite flakes after mechanical exfoliation (left panel) and after thermal annealing in Ar/H$_2$ mixture in a tubular furnace at 300°C (right panel).*